\documentclass[12pt,showpacs,amsmath,amssymb]{iopart}

\usepackage[utf8x]{inputenc}
\begin{document}

\title[Gaussian Potential]{Asymptotic Iteration and Variational Methods for Gaussian Potential}

\author{Halil Mutuk}

\address{Physics Department, Faculty of Arts and Sciences, Ondokuz Mayis University, 55139, Samsun, Turkey}
\ead{halilmutuk@gmail.com}
\vspace{10pt}

\begin{abstract}
In this paper we studied approximate solutions of the radial Schrödinger equation with the attractive Gaussian potential. We used asymptotic iteration method and variational method in order to obtain energy eigenvalues for any $n$ and $l$ quantum numbers. Our results are in good agreement with the other studies. 
\end{abstract}


\vspace{2pc}
\noindent{\it Keywords}: Gaussian Potential, Asymptotic Iteration Method, Variational Method
%\maketitle

\section{Introduction}
Schr\"{o}dinger equation of a particle with mass $m$ and potential $V$ can be written as
\begin{equation}
-\frac{\hbar^2}{2m}\nabla^2\psi (\vec{r})+V(\vec{r})\psi (\vec{r})=E\psi (\vec{r}).
\end{equation}   
Here $\psi (\vec{r})$ is the wave function and $E$ is the energy eigenvalue.
Since its invention at the beginning of 1900s, Schrödinger equation  is applied to electrons, protons, neutrons, quarks and etc. Solutions of Schrödinger equation is an important issue in physics, mathematics, chemistry and also in biology.   

The solution of Schrödinger equation have been always a challenge for most of the time. Many attempts have been made to solve Schrödinger equation analytically or approximately. The solutions of Schrödinger equation is important to understand energy spectrum and system evolution in time (in case of time-dependent Schrödinger equation). If one gets energy and wave function for the related system, it is possible to have a description about that system. The general framework is to solve Schrödinger equation with a potential of $V(r)$. In most cases it is not always possible to obtain an accurate solution. For example the Morse potential \cite{1} , $V(r)=D_e \left[ e^{-2a(r-r_e)-2e^{a(r-r_e}} \right]$, where $D_e$ is the association energy, $r_e$ is the equilibrium distance and $a$ is the potential parameter, has no exact solutions. In that cases approximate solutions are seeking to describe the system. Beside that some potentials are exactly solvable within the framework of Schrödinger equation such as harmonic oscillator potential, Coulomb potenital, Morse potential \cite{2}, Eckart potential \cite{3}, Woods-Saxon potential \cite{4} and Kratzer potential \cite{5}.   

The general method to solve Schrödinger equation is to transform it into a well-known ordinary differential equation which have solutions in terms of Hermit, Legendre, Bessel or in any other form of hypergeometric functions. These are generally appear in undergraduate/graduate quantum mechanics textbooks such as \cite{6,7,8,9}. An alternative method is Supersymmetric Quantum Mechanics (SUSY-QM). In SUSY-QM, the Hamiltonian of the system is factorized. The other method is Nikiforov-Uvarov in which one can transform Schrödinger equation with a specific potential into a  hypergeometric-type second order differential equation. This method can be desrcibed as by-hand method since one can see all the steps in the solution. Perturbation and variational methods can be also used to find solutions. Of course there is no one way to solve eigenvalue equations like Schrödinger equation. There is a rich literature on that subject. 

In this paper we will seek solutions of Gaussian potential by asymptotic iteration and variational methods. Variational methods for Gaussian potential have been studied before with different wave functions but to the best of our knowledge, a study of asymptotic iteration method for Gaussian potential has not been reported.  In section 2, we will introduce Gaussian potential. In sections 3 and 4 asymptotic iteration and variational methods are given, respectively. In the last section these methods are applied to Gaussian potential and we present our results. 

\section{Gaussian Potential} 

Gaussian potential is in the form of 
\begin{equation}
V(r)=-Ae^{-\lambda r^2},
\end{equation}
where $A$ is the depth of the potential and $\lambda$ is a parameter. It is an attractive potential. Gaussian potential was used as a model in nucleon-nucleon scattering theory by Buck et al. \cite{10}. The eigenvalues of Gaussian potential were first obtained by Buck in a unpublished paper at 1977 via direct numerical integration. Stephenson \cite{11} obtained eigenvalues of the three-dimensional Schrödinger equation with a radial Gaussian potential by using the Liouville-Green uniform asymptotic method. Bessis et al. \cite{12} studied  energies and eigenfunctions of the Schrödinger equation with a radial gaussian potential by  using a perturbational and variational treatment on a conveniently chosen basis of transformed Jacobi functions. Lai obtained energy eigenvalues of the gaussian potential for various eigenstates within the framework of the hypervirial-Pade scheme \cite{13}. Cohen obtained eigenvalues and approximate eigenfunctions from a first-order perturbation treatment based on a scaled harmonic oscillator model \cite{14}. $1/N$ expansion was applied to Gaussian potential by Chatterjee \cite{15}. Koksal in \cite{16} used a combination of perturbation theory and supersymmetric quantum theory for bound state energies of an electron confined in an attractive Gaussian potential. 

This work is devoted to obtain eigenvalues of the radial Schrödinger equation with the Gaussian potential

\begin{equation}
\frac{d^2 R_{nl}(r)}{dr^2}+\frac{2m}{\hbar^2}\left( E-V(r)-\frac{l(l+1)}{2mr^2} \right) R_{nl}(r)=0, \label{eqn1}
\end{equation}
 where $m$ is the mass of the particle. Inserting Gaussian potential in Eqn. (\ref{eqn1}) radial Schrödinger equation takes the following form:
 
 \begin{equation}
\frac{d^2 R_{nl}(r)}{dr^2}+\frac{2m}{\hbar^2}\left( E+Ae^{-r^2}-\frac{l(l+1)}{2mr^2} \right) R_{nl}(r)=0, \label{eqn2}
\end{equation}
where we took $\lambda=1$. This form of Schrödinger equation cannot be solved analytically.

\section{Asymptotic Iteration Method}
The asymptotic iteration method was introduced to  solve second-order homogeneous linear differential equations of the form

\begin{equation}
y^{\prime \prime}(x)=\lambda_0(x) y^\prime (x) + s_0(x) y_0(x) \label{eqn3}
\end{equation}

where $\lambda_0(x)\neq 0$ and the variables $\lambda_0(x)$ and $s_0(x)$ have sufficiently many derivatives \cite{17}. This differential equation has a solution as follows:

\begin{equation}
y(x)=\texttt{exp} \left( -\int^x \alpha dt \right) \left[ C_2+C_1 \int^x \texttt{exp} (\int^t \left[(\lambda_0(\tau)+2\alpha(\tau))d\tau \right]   dt  \right] \label{eqn4}
\end{equation}
 where 
\begin{equation}
\frac{s_k(x)}{\lambda_k(x)}=\frac{s_{k-1}(x)}{\lambda_{k-1}(x)} \equiv \alpha(x). \label{eqn5}
\end{equation}
for sufficiently large $k$.  In Eqn. (\ref{eqn5}) $s_k(x)$ and $\lambda_k(x)$ are defined as follows:

\begin{eqnarray}
\lambda_k(x)=\lambda^\prime_{k-1}(x)+s_{k-1}(x)+\lambda_0(x) \lambda_{k-1}(x), \nonumber \\
s_n(x)=s^\prime_{k-1}(x)+s_0(x)\lambda_{k-1}(x),~ k=1,2,3, \cdots \label{eqn6}
\end{eqnarray}

The convergence (quantization) condition can be defined as 
\begin{equation}
\delta_k(x)=\lambda_k(x)s_{k-1}(x)-\lambda_{k-1}(x)s_k(x)=0, ~ k=1,2,3, \cdots  \label{eqn10}
\end{equation}
For a given radial potential, it is possible to convert radial Schrödinger equation into Eqn. (\ref{eqn3}). Once this form has been obtained, it is easy to see $s_0(x)$ and $\lambda_0(x)$ and calculate  $s_k(x)$ and $\lambda_k(x)$ by using Eqn. (\ref{eqn6}). Eigenvalues are obtained from the quantization (convergence) condition, $\delta_k(x)=0$. 

\subsection{AIM for Gaussian Potential}

We have said that Eqn. (\ref{eqn2}) cannot be solved exactly. One way to solve this differential equation is to use Taylor expansion. If we expand Gaussian potential near origin we get
\begin{equation}
V(r)=-A+A r^2-\frac{A r^4}{2}+\frac{A r^6}{6}-\frac{A r^8}{24}+\frac{A r^{10}}{120}+O\left(r^{11}\right). \label{eqn7}
\end{equation} 

In principle one can truncate this series at higher powers. But since it is an asymptotic expansion, the major contributions come from the first few terms.  Putting this expansion into the radial Schrödinger equation and defining $\epsilon=\frac{2mE_n}{\hbar^2}$ and $\tilde{A}=\frac{2mA}{\hbar^2}$ one can get

\begin{equation}
\frac{d^2 R_{nl}(r)}{dr^2} + \left\lbrace \epsilon -\tilde{A} \left[  1- r^2 +\frac{r^4}{2}-\frac{r^6}{6}+\frac{r^8}{24}-\frac{r^{10}}{20}\right]+\frac{l(l+1)}{r^2} \right\rbrace R_{nl}(r)=0 \label{eqn8}.
\end{equation}
 
This is a second order linear differential equation and in order to solve this equation via AIM, we should transform it into Eqn. (\ref{eqn3}). In a work of Karakoc and Boztosun  \cite{18}, they solved Schrödinger equation with Yukawa potential $V(r)=-\frac{A}{r} \texttt{exp} (-\alpha r)$ via AIM by suggesting a wave function of the form $R_{nl}(r)=r \texttt{exp} (-\beta r)f_{nl}(r)$. In the present work, regarding the shape of the potential we propose a wave function of the form

\begin{equation}
R_{nl}(r)=r^{l+1} \texttt{exp}(-\beta r^2)f_{nl}(r) \label{eqn11}.
\end{equation}

Inserting this wave function in Eqn. (\ref{eqn8}) we have second order linear homogeneous differential equation as

\begin{eqnarray}
\frac{d^2 f_{nl}(r)}{dr^2}&=&\left( -\frac{2 (L+1)}{r} + 4 \beta  r \right) \frac{df_{nl}(r)}{dr} \nonumber \\  &+& [\frac{1}{120} A \left(r^{10}-5 r^8+20 r^6-60 r^4+120 r^2-120\right) \nonumber \\ &-&\epsilon +2 \beta  \left( 2 L-2 \beta  r^2+3 \right) ]  f_{nl}(r) \label{eqn9}.
\end{eqnarray}

By comparing Eqn. (\ref{eqn9}) with Eqn. (\ref{eqn3}), we see that  AIM can be applied and $\lambda_0(x)$ and $s_0(x)$ can be written as follows:
\begin{eqnarray}
\lambda_0(x)=\left( -\frac{2 (L+1)}{r} + 4 \beta  r \right) \nonumber \\
s_0(x)=[\frac{1}{120} A \left(r^{10}-5 r^8+20 r^6-60 r^4+120 r^2-120\right) \nonumber \\ - \epsilon +2 \beta  \left( 2 L-2 \beta  r^2+3 \right) ]
\end{eqnarray}

 At this point, we should mention an important concept about AIM. If the potential is an exactly solvable around $x_0$, the solution can be obtained when the iteration number, $k$ equals the principal quantum number (radial quantum number) $n$. If the potential is not exactly solvable, for iteration number $k$ bigger than radial quantum number $n$, a suitable $x$ is chosen and energy eigenvalues can be obtained \cite{17}.  Energy eigenvalues  can be obtained by the quantization condition Eqn. (\ref{eqn10}). The quantization condition depends on $x$ but the energy eigenvalues obtained from this condition should be independent of the choice of $x$. This suitable choice of $x$ minimizes the potential or maximizes the wave function, Eqn. (\ref{eqn11}). In the AIM method, the wave function can be written as
 \begin{equation}
 R(x)=f(x)g(x)
 \end{equation}
where $f(x)$ represents asymptotic behaviour. In our case, $f(x)=x^{l+1} \texttt{exp}(-\beta x^2)$ and $x=\frac{\sqrt{l+1}}{\sqrt{2} \sqrt{\beta}}$. $\beta$ is an arbitrary parameter related to the convergence \cite{19}. In Table \ref{jlab1}, the convergence for different values of $\beta$ is shown.

\begin{table}
\caption{\label{jlab1} The convergence of the eigenvalues for different $\beta$ values with $n=0$ and $l=0$. $k$ is the iteration number.}
\normalsize
\begin{tabular}{@{}llllll}
\br
$k$ & $\beta=5$ & $\beta=10$ & $\beta=15$ & $\beta=20$ & $\beta=25$\\
\mr
5  & 341.836822674 & 341.863607546 & 340.893507947 & 337.720033882  & 332.697776132 \\
10 & 341.895185264 & 341.894773908 & 341.840149675 & 341.367656610  & 340.065208776  \\
15 & 341.895181270 & 341.895177621 & 341.891722431 & 341.825947031  & 341.515599948  \\
20 & 341.895182710 & 341.895183501 & 341.894944170 & 341.886006151  & 341.817565584 \\
25 & 341.895182759 & 341.894765159 & 341.895345639 & 341.897695810  & 341.870927989 \\
30 & 341.895185517 & 341.910303257 & 342.154549418 & 342.727071834  & 355.888111258 \\
35 & 341.895475114 & 341.910303265 & 341.894010908 & 342.733072138  & 353.522636070 \\
\br
\end{tabular}\\
\end{table}
\normalsize

As can be seen in Table \ref{jlab1}, $\beta=5$ and $\beta=10$ gave more sensical results. The energy eigenvalues in this paper are obtained with $\beta=10$ since the solutions are very close after two iterations. The results are shown in Table \ref{jlab2}. 

\section{Variational Method}
 
Variational method, known as Rayleigh-Ritz method, is a very useful in obtaining energy eigenvalues and eigenstates of the related system. This method is used when perturbation theory cannot be applied, i.e. Hamiltonian cannot be written simply $H=H_0+V$ where $H_0$ is the Hamiltonian which can be solved exactly. In this method, we have a system that is solvable and have energy eigenvalue and eigenfunction of one state, generally ground state. By having these, one can get energy eigenvalues and eigenfunctions of higher states. 

The Hamiltonian of the Gaussian potential is
\begin{equation}
H=\frac{p^2}{2\mu}+V(r)=\frac{p^2}{2\mu}-Ae^{- r^2},
\end{equation}
where $p$ is the relative momentum, $\mu$ is the reduced mass of the two particles, $\mu=\frac{m_1 m_2}{m_1+m_2}$. In our case $m_1=m_2$. In order to obtain energy eigenvalues, we have solved radial Schrödinger equation with the Hamiltonian above using the variational method
\begin{equation}
E=\frac{\langle R(r) \vert H \vert  R(r) \rangle}{\langle \psi \vert \psi \rangle} \equiv \langle H \rangle.
\end{equation}
We used radial part of three dimensional harmonic oscillator wave function as trial function
\begin{equation}
R_{nl}=N(\frac{r}{b})^l L_n^{l+\frac{1}{2}}(\frac{r^2}{b^2})e^{-\frac{r^2}{b^2}} \label{eqn12}
\end{equation} 
with the $L_n^{l+\frac{1}{2}}(x)$ is the associated Laguerre polynomials and normalization constant 
\begin{equation}
\vert N \vert^2=\frac{2n!}{b^3 \sqrt{\pi}} \frac{2^{2n+2l+1}}{(2n+2l+1)!}(n+l)!. 
\end{equation}
In Eqn. (\ref{eqn12}), $b$ is treated as the variational parameter which is determined  by minimizing the expectation value of the Hamiltonian for each state. After having $b$ value, it was used in the wave function to obtain energy eigenvalues. The results of variational method are given in Table \ref{jlab2}.

\begin{table}[h]
\caption{\label{jlab2} Energy eigenvalues of the Gaussian potential in units $\hbar=1$, $m=0.5$.}
\footnotesize
\begin{tabular}{@{}lllllllll}
\br
n&l& AIM & Variational &  \cite{10}  & \cite{11} &  \cite{12}$^{a}$  & \cite{13} & \cite{15}$^{b}$ \\
\mr
0& 0 & 341.895 & 341.895 & 341.9 & 341.6 & 341.895 & 341.895 & 341.895 \\
& 1 & 304.464 & 304.748 & 304.5 & 304.2 & 304.463 & 304.463 & 304.463 \\
& 2 & 268.111 & 268.235 & 268.1 & 267.9 & 268.111 & 268.111 & 268.111 \\
& 3 & 232.873 & 233.090 & 232.9 & 232.6 & 232.875 & 232.875 & 232.875 \\
& 4 & 198.796 & 198.925 & 198.8 & 198.5 & 198.797 & 198.798 & 198.798 \\
& 5 & 165.982 & 166.174 & 165.9 & 165.7 & 165.925 & 165.928 & 165.928 \\
& 6 & 134.321 & 134.450 & 134.3 & 134.1 & 134.317 & 134.323 & 134.323 \\
& 7 & 104.121 & 104.269 & 104.1 & 103.9 & 104.040 & 104.051 & 104.051 \\
\hline
1& 0 & 269.643 & 269.644 & 269.7 & 269.4 & 269.644 & 269.644 & 269.644 \\
 & 1 & 235.469 & 235.696 & 235.5 & 235.2 & 235.450 & 235.450 & 235.451 \\
 & 2 & 202.415 & 202.592 & 202.4 & 202.2 & 202.431 & 202.431 & 202.432 \\
 & 3 & 170.566 & 170.933 & 170.4 & 170.6 & 170.639 & 170.639 & 170.640 \\ 
 & 4 & 140.028 & 140.249 & 140.1 & 139.9 & 140.134 & 140.135 & 140.136 \\ 
 & 5 & 110.745 & 111.105 & 111.0 & 110.7 & 110.989 & 110.993 & 110.994 \\ 
 & 6 & 83.359 & 83.426 & 83.3 & 83.0 & 83.295 & 83.306 & 83.307 \\ 
 & 7 & 56.467 & 57.315 & 57.2 & 56.9 & 57.171 & 57.196 & 87.197 \\ 
 \hline
2& 0 & 203.958 & 203.983 & 204.0 & 203.7 & 203.983 & 203.983 & 203.997 \\
 & 1 & 173.222 & 173.483 & 173.3 & 173.0 & 173.244 & 173.244 & 173.257 \\
 & 2 & 143.669 & 143.923 & 143.8 & 143.6 & 143.808 & 143.809 & 143.821 \\
 & 3 & 115.588 & 115.869 & 115.8 & 115.5 & 115.752 & 115.754 & 115.765 \\ 
 & 4 & 88.992  & 89.281 & 89.2 & 88.9 & 89.169 & 89.175 & 89.184 \\
 & 5 & 63.077 & 64.305 & 64.2 & 63.9 & 64.180 & 64.196 & 64.202 \\
 & 6 & 39.874 & 40.169 & 41.0 & 40.7 & 40.950 & 40.989 & 40.990 \\
 & 7 & 18.964 & 19.508 & 19.8 & 19.5 & 19.721 & 19.813 & 19.809 \\
 \hline
3& 0 & 145.133 & 145.377 & 145.4 & 145.1 & 145.377 & 145.378 & 145.431 \\ 
 & 1 & 118.718 & 118.636 & 118.4 & 118.1 & 118.381 & 118.384 & 118.432 \\ 
 & 2 & 92.089 & 92.733 & 92.9 & 92.6 & 92.873 & 92.878 & 92.919 \\ 
 & 3 & 67.624 & 68.843 & 69.0 & 68.7 & 68.973 & 68.984 & 69.013 \\ 
 & 4 & 46.950 & 46.692 & 46.9 & 46.6 & 46.846 & 46.868 & 46.881 \\  
 & 5 & 26.736 & 26.594 & 26.8 & 26.5 & 26.727 & 26.778 & 26.768 \\   
 & 6 & 9.667 & 8.953 & 9.1 & 8.8 & 8.995 & 9.126 & 9.080 \\ 
 \hline  
4& 0 & 93.497 & 94.456 & 94.5 & 94.2 & 94.454 & 94.458 & 94.582 \\ 
 & 1 & 71.881 & 71.904 & 71.6 & 71.4 & 71.617 & 71.624 & 71.720 \\  
 & 2 & 49.021 & 50.842 & 50.6 & 50.3 & 50.557 & 50.568 & 50.623 \\ 
 & 3 & 31.536 & 31.793 & 31.5 & 31.3 & 31.499 & 31.521 & 31.518 \\  
 & 4 & 14.252 & 14.042 & 14.9 & 14.6 & 14.794 & 14.851 & 14.768 \\  
 & 5 & 1.182 & 1.369 &  &  & 1.1 & 1.295 & 1.030 \\  
 
\br
\end{tabular}\\
$^{a}$ Variational
$^{b}$ Shifted $1/N$ expansion
\end{table}

\section{Results and Discussion}
In this paper, we solved radial Schrödinger equation with the radial Gaussian potential via Asymptotic iteration and variational methods. Gaussian potential was studied before via variational, perturbation, hypervirial-Pade and $1/N$ expansion methods mentioned before as well as direct numerical integration in \cite{20}. 

Bessis et al. studied Gaussian potential with square integrable eigenfunctions in terms of Jacobi polynomial \cite{12}. In present work we used rather a different wave function in terms of Laguerre polynomials. The variational method is a cumbersome method to obtain energy eigenvalues and eigenfunctions. It contains tedious calculations. If we don't know exact results of a given Hamiltonian, variational method can be used. We choose a wave function which mimickes the actual unknown wave function of the system. This wave function can depend on some parameters. In some cases trial wave function can be normalized to 1 so that numbers of unknown parameters can be decreased. The expectation value of the Hamiltonian is  calculated. The expectation value of the Hamiltonian must be bigger than the ground state, $\langle H \rangle > E_0$ where $E_0$ is the ground state energy of the system which is exactly solvable. There is no an {\it a priori} reason for the choice of trial wave function. One must be careful about the behaviour of the wave function at asymptotic ranges. 

We have solved the radial Gaussian potential for any $n$, $l$ and $A$ in the framework of AIM by transforming Schrödinger equation into a second-order differential equation in the form of Eqn. (\ref{eqn3}). In AIM, there is no constraint on the potential parameter values which makes AIM easy to implement. 

The obtained energy eigenvalues by AIM and variational methods are in good agreement with the other results given in Table \ref{jlab2}. 

\section{Conclusion}
Schrödinger equation which is an eigenvalue equation can be solved by analytically and approximately.  Interactions under a potential function generally make Schrödinger equation solvable only by approximation methods. For example WKB method, Shifted 1/N exansion method, Pade method and Nikiforov-Uvaranov method are used to solve Schrödinger equation approximately.

Asymptotic iteration method deals with the second order linear homogeneous differential equations. If one have a wave function with appropriate boundary conditions, AIM can be applied to such problems. In this manner eigenvalues can be obtained with a good accuracy. The AIM can be extended to study other interacting potentials together with the other methods.

\end{document}